\documentclass{sasar}
\usepackage{txfonts}
\usepackage{multirow}
\usepackage{hyperref}
\usepackage{ragged2e}
\usepackage{multicol}
\usepackage{float}

\def\currentvolume{2}
\def\currentissue{1}
\def\currentyear{2026} 
\def\ppages{161-179} 
\def\DOI{10.64389/mjs.2026.02145}
\numberwithin{equation}{section}

\makeatletter
\renewcommand{\@biblabel}[1]{#1\hfill \hspace{-0.2cm}}
\makeatother

\providecommand{\typeofarticle}{Research article}
\providecommand{\affil}[1]{\textsuperscript{#1}}
\providecommand{\addr}[2]{\textsuperscript{#1}#2}
\providecommand{\corraddr}[1]{\textsuperscript{*}#1}

\begin{document}

\title{Modi linear failure rate distribution with application to survival time data}

\author{Lazhar Benkhelifa\affil{1,*}}

\address{\small
	\addr{\affilnum{1}}{Department of Mathematics, Mohamed Khider University, Biskra, Algeria}
}
\corraddr{lazhar.benkhelifa@univ-biskra.dz}

\maketitle
\vspace{-5pt}

\noindent
\begin{minipage}[t]{0.33\textwidth}
    \raggedright
    \textbf{ARTICLE INFO}

    \vspace{1mm}
    {\footnotesize 
   Linear failure rate distribution\\
   moment\\
   order statistic \\
    maximum likelihood

    \vspace{2mm}

    \vspace{2mm}
\textbf{Important Dates:} \\
Received: 5 November 2025 \\
Revised: 26 December 2025 \\
Accepted: 29 January 2026 \\
Online: 1 February 2026 \\
}

\vspace{-2mm} 
\begin{flushleft}
   \href{https://creativecommons.org/licenses/by/4.0/#CC}{%
\includegraphics[scale=0.35]{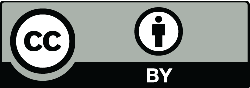}%
}
\end{flushleft}
\vspace{-5mm}
{\footnotesize
\href{https://creativecommons.org/licenses/by/4.0/}{Copyright © \currentyear}
by the authors. Published under Creative Commons Attribution (CC BY) license.}
\vspace{40pt}

\end{minipage}
\hfill
\begin{minipage}[t]{0.64\textwidth}
\justifying

   \noindent \textbf{ABSTRACT}

    \vspace{1mm}
A new lifetime model, named the Modi linear failure rate distribution, is suggested. This flexible model is capable of accommodating a wide range of hazard rate shapes, including decreasing, increasing, bathtub, upside-down bathtub, and modified bathtub forms, making it particularly suitable for modeling diverse survival and reliability data. Our proposed model contains the Modi exponential distribution and the Modi Rayleigh distribution as sub-models. Numerous mathematical and reliability properties are derived, including the $r^{th}$ moment, moment generating function, $r^{th}$ conditional moment, quantile function, order statistics, mean deviations, Rényi entropy, and reliability function. The method of maximum likelihood is employed to estimate the model parameters. Monte Carlo simulations are presented to examine how these estimators perform. The superior fit of our newly introduced model is proved through two real-world survival data sets.
\end{minipage}

\begin{center}
\vspace{-20pt}
\rule{\textwidth}{0.4pt}
\end{center}

\section{Introduction}\label{s1}
    Despite the availability of the thousands of lifetime models in reliability, survival analysis, and related domains, the quest for a more flexible distribution persists to this day. This ongoing search is fueled by the complexity and variety of real-world data sets, which often exhibit non-monotonic hazard rates, heavy tails, skewness, or heterogeneity. For this reason, several methods have been proposed to generate new distributions by adding one or more extra shape parameters to the baseline model. For example, the power generalization method was used by \cite{GuptaKundu2001} to introduce the exponentiated exponential model. The beta-G distribution was defined by \cite{EugeneLeeFamoye2002}. The DUS transformed method was developed by \cite{KumarSinghSingh2015} and used the exponential distribution as the parent model. Mahdavi and Kundu \cite{MahdaviKundu2017} developed the alpha power transformation technique. Benkhelifa \cite{Benkhelifa2022} developed the alpha power Topp-Leone-G and used the Weibull distribution as the baseline model. Kavya and Manoharan \cite{KavyaManoharan2021} suggested the Kavya-Manoharan transformation technique. Modi et al. \cite{ModiKumarSingh2020} introduced an interesting method of proposing new distributions called the Modi family of distributions and used the exponential distribution as the baseline model. Making use of the Modi family, some authors introduced the novel distribution and provided its properties. We mention: Kumawat et al. \cite{KumawatModiNagar2023} introduced the Modi Weibull distribution. Muhimpundu et al. \cite{MuhimpunduOdongoKube2025} suggested the Modi exponentiated inverted Weibull distribution. Akhila and Girish Babu \cite{AkhilaBabu2025} presented the Modi Fréchet distribution. Kumar et al.\cite{KumarMeenaShukla2025} introduced the Modi Rayleigh distribution.  
    The linear failure rate (LFR) distribution, or the linear exponential distribution, which has the Rayleigh and exponential distributions as sub-models, is widely used in reliability engineering and survival analysis. For example, Carbone et al. \cite{CarboneKellerhouseGehan1967} applied the LFR distribution to model the survival patterns of patients with plasmacytic myeloma. Despite its usefulness in scenarios with monotonic increasing failure rates, the LFR distribution is inadequate to model the data that exhibit non-linear or non-monotonic hazard rates, such as those with bathtub-shaped or unimodal failure rates. The beta LFR distribution ( \cite{JafariMahmoudi2015}), the Harris generalized linear exponential distribution (\cite{albin2021harris}), the modified beta linear exponential distribution (\cite{BakouchSaboorKhan2021}, the transmuted generalized linear exponential distribution (\cite{GhoshKatariaVellaisamy2021}), the Weibull linear exponential distribution (\cite{AtiaMahmoudSagheerDesouky2023} and the weighted LFR distribution (Wang e In this article, we introduce a new extension of the LFR distribution by employing the Modi family of distributions, termed Modi LFR (MLFR) distribution.  
    Our article is organized as follows. Section \ref{s2} introduces the MLFR distribution. The mathematical and reliability properties of the proposed model are derived in Section \ref{sec3}. The maximum likelihood estimators of the MLFR distribution parameters are discussed in Section \ref{sec4}, and the performance of these estimators is examined through a simulation study in Section \ref{sec5}. Two survival time data sets are analyzed in Section \ref{sec6} to demonstrate the flexibility of the new model. Concluding remarks and future work are provided in Section \ref{sec7}.
    
\section{The MLFR distribution}\label{s2}
The Modi family of distributions was developed by \cite{ModiKumarSingh2020}. The Modi family is characterized by the following cumulative distribution function (CDF):

\begin{equation}\label{eq1}
F(x) = \frac{(1 + \alpha^{\beta})G(x)}{\alpha^{\beta} + G(x)}, \quad x, \alpha, \beta > 0
\end{equation}

where $G(x)$ is the baseline CDF. The CDF and the probability density function (PDF) of the LFR distribution with parameters $a$ and $b$ are given respectively by

\begin{equation} \label{eq:2}
G(x)=1-e^{-ax-(b/2)x^2},
\end{equation}

and

\begin{equation*}
g(x)=(a+bx)e^{-ax-(b/2)x^2},
\end{equation*}

where $x\geq 0$, $a\geq 0$ and $b\geq 0$ with $a+b>0$.

Therefore, the CDF of the MLFR distribution is obtained by replacing (\ref{eq:2}) in (\ref{eq1}):

\begin{equation} \label{eq:3}
F(x)=\frac{(1+\alpha^{\beta})(1-e^{-ax-(b/2)x^2})}{\alpha^{\beta}+1-e^{-ax-(b/2)x^2}},
\end{equation}

and the PDF of the MLFR distribution is

\begin{equation} \label{eq:4}
f(x)=\frac{\alpha^{\beta}(a+bx)e^{-ax-(b/2)x^2}}{(1+\alpha^{\beta})\left(1-\frac{e^{-ax-(b/2)x^2}}{1+\alpha^{\beta}}\right)^2}.
\end{equation}

The reliability or survival function of MLFR distribution is

\begin{equation*}
S(x)=\frac{\alpha^{\beta}}{(\alpha^{\beta}+1)e^{ax+(b/2)x^2}-1},
\end{equation*}

whereas, the hazard rate function of MLFR distribution is

\begin{equation*}
h(x)=\frac{(\alpha^{\beta}+1)(a+bx)}{\alpha^{\beta}+1-e^{-ax-(b/2)x^2}}.
\end{equation*}

It is clear that when $b=0$, we get the Modi exponential distribution which is introduced by Modi et al. \cite{ModiKumarSingh2020}, whereas when $a=0$, we get the Modi Rayleigh distribution which is proposed by   \cite{KumarMeenaShukla2025}.

\begin{figure}[ht!]
\centering
\includegraphics[scale=0.65]{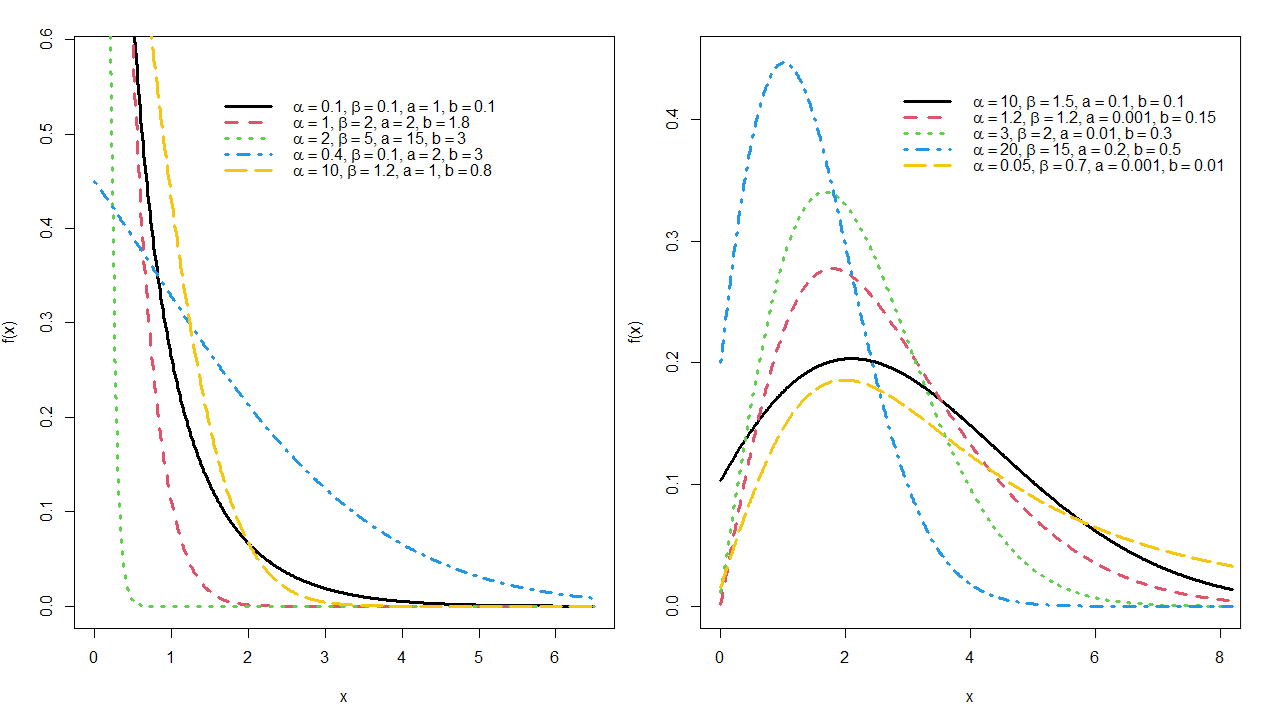}
\caption{PDF plot for $\beta$, $\alpha$, $a$, and $b$}
\label{fig1}
\end{figure}

\begin{figure}[ht!]
\centering
\includegraphics[scale=0.75]{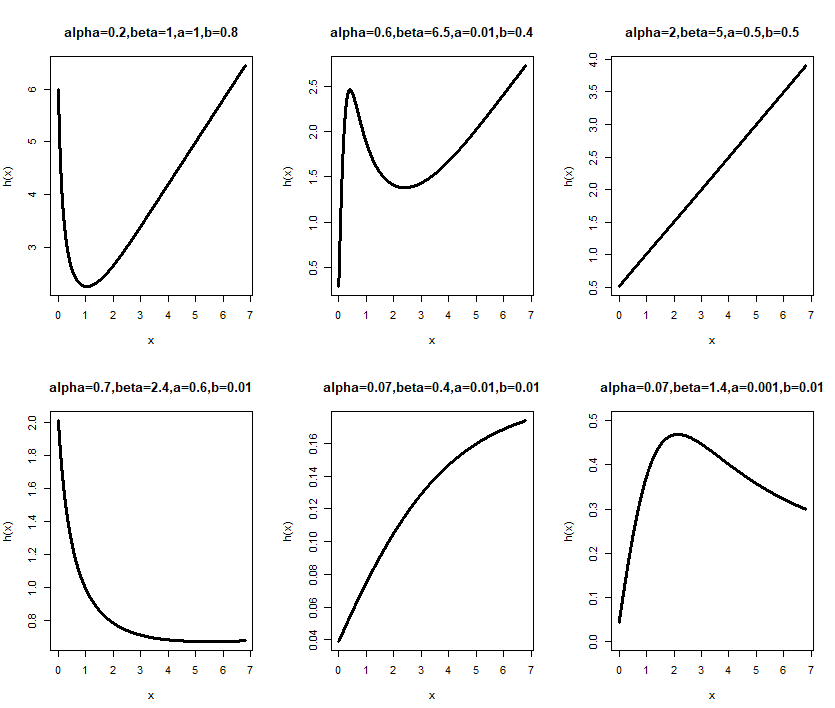}
\caption{Hazard rate plot for $\beta$, $\alpha$, $a$, and $b$}
\label{fig2}
\end{figure}

For selected values of $\beta$, $\alpha$, $a$, and $b$, the PDF plots and the hazard rate function for the MLFR distribution are shown in Figures \ref{fig1} and \ref{fig2}, respectively. It is evident from Figure \ref{fig1} that the PDF of the MLFR distribution is unimodal and can be decreasing or non-monotone. Figure \ref{fig2} reveals that the hazard rate can be decreasing, increasing, upside-down, bathtub shaped, or modified bathtub shaped (unimodal shape followed by increasing). This indicates that the MLFR distribution is highly flexible and well-suited for modeling diverse types of lifetime data.

\section{Mathematical and Reliability Properties}\label{sec3}

Some properties of our proposed MLFR distribution are presented in this part.

\subsection{Moments}

The $r$th moment about the origin of the MLFR distribution is

\begin{equation*}
\mu_{r}' = \int_{0}^{\infty} x^{r} \frac{\alpha^{\beta}(a+bx)e^{-ax-(b/2)x^{2}}}{(1+\alpha^{\beta})\left(1-\frac{e^{-ax-(b/2)x^{2}}}{1+\alpha^{\beta}}\right)^{2}} dx.
\end{equation*}

Making use of Equation (5.2.11.3), in \cite{PrudnikovBrychkovMarichev1986}, which is

\begin{equation}\label{eq5}
\frac{s!}{(1-v)^{s+1}} = \sum_{k=0}^{\infty} \frac{(s+k)!}{k!} v^{k},
\end{equation}

we obtain

\begin{equation*}
\mu_{r}' = \frac{\alpha^{\beta}}{1+\alpha^{\beta}} \sum_{k=0}^{\infty} \frac{k+1}{(1+\alpha^{\beta})^{k}} \int_{0}^{\infty} (ax^{r}+bx^{r+1})\left(e^{-ax-(b/2)x^{2}}\right)^{k+1} dx.
\end{equation*}

From \cite{BakouchSaboorKhan2021}, we have

\begin{equation}\label{eq6}
\int_{0}^{\infty} x^{s} e^{-\beta x^{\gamma}} e^{-ax} dx = \frac{1}{\gamma \beta^{(s+1)/\gamma}} \sum_{m=0}^{\infty} \frac{(-1)^{m}}{m!} \left(\frac{\alpha}{\beta^{1/\gamma}}\right)^{m} \Gamma\left(\frac{s+m+1}{\gamma}\right),
\end{equation}

where $\Gamma(\cdot)$ denotes the gamma function. Hence, after some algebra, we get

\begin{equation*}
\mu_{r}' = \sum_{k=1}^{\infty} \sum_{m=1}^{\infty} \frac{(-ak)^{m-1} \alpha^{\beta} k}{(m-1)!(1+\alpha^{\beta})^{k}} \left\{ \frac{a \Gamma\left(\frac{r+m}{2}\right)}{2(kb/2)^{((r+m)/2)}} + \frac{b \Gamma\left(\frac{r+m+1}{2}\right)}{2(kb/2)^{((r+m+1)/2)}} \right\}.
\end{equation*}

\subsection{Moment Generating and Characteristic Functions}

The moment generating function of the MLFR distribution is

\begin{equation*}
M_{X}(t) = E(e^{tX}) = \int_{0}^{+\infty} e^{tx} \frac{\alpha^{\beta}(a+bx)e^{-ax-(b/2)x^{2}}}{(1+\alpha^{\beta})\left(1-\frac{e^{-ax-(b/2)x^{2}}}{(1+\alpha^{\beta})}\right)^{2}} dx.
\end{equation*}

Using \eqref{eq5} and \eqref{eq6}, and after some simplifications, we obtain

\begin{equation*}
M_{X}(t) = \sum_{k=1}^{\infty} \sum_{m=0}^{\infty} \frac{(t-ka)^{m} \alpha^{\beta} k}{m! (1+\alpha^{\beta})^{k} (kb/2)^{m/2}} \left\{ \frac{a \Gamma\left(\frac{m+1}{2}\right)}{\sqrt{2bk}} + \frac{\Gamma\left(\frac{m+2}{2}\right)}{k} \right\}.
\end{equation*}

Similarly, one can obtain the characteristic function as

\begin{equation*}
\varphi_{X}(t) = E(e^{itX}) = \sum_{k=1}^{\infty} \sum_{m=0}^{\infty} \frac{(it-ka)^{m} \alpha^{\beta} k}{m! (1+\alpha^{\beta})^{k} (kb/2)^{m/2}} \left\{ \frac{a \Gamma\left(\frac{m+1}{2}\right)}{\sqrt{2bk}} + \frac{\Gamma\left(\frac{m+2}{2}\right)}{k} \right\}, \quad i=\sqrt{-1}.
\end{equation*}

\subsection{Quantile Function}

The quantile function $Q(u)$ of the MLFR distributions is the inverse of $F$ which given in (\ref{eq:3} ). After some algebra, we get

\begin{equation}\label{eq7}
Q(u) = \frac{1}{b} \left( -a + \sqrt{a^{2} + 2b \log\left( \frac{u-1-\alpha^{\beta}}{(1+\alpha^{\beta})(u-1)} \right)} \right), \quad \text{where } u \in (0,1).
\end{equation}

The median $M$ of the MLFR distribution is derived by replacing $u=0.5$ in \eqref{eq7} while the first and third quartiles are obtained by replacing $u=0.25$ and $u=0.75$ in \eqref{eq7}, respectively. For example, we have

\begin{equation*}
M = Q\left(\frac{1}{2}\right) = \frac{1}{b} \left( -a + \sqrt{a^{2} + 2b \log\left( \frac{2\alpha^{\beta}+1}{\alpha^{\beta}+1} \right)} \right).
\end{equation*}

\subsection{Conditional Moments}

For lifetime models, the $r$th conditional moment $E(X^{r}|X>t)$, plays a pivotal role in prediction. The $r$th conditional moment of the MLFR distribution is

\begin{equation*}
E(X^{r}|X>t) = \frac{1}{S(x)} \int_{t}^{\infty} x^{r} f(x) dx = \frac{(\alpha^{\beta}+1)e^{ax+(b/2)x^{2}}-1}{\alpha^{\beta}} \int_{t}^{\infty} x^{r} f(x) dx,
\end{equation*}

where

\begin{equation*}
\int_{t}^{\infty} x^{r} f(x) dx = \frac{\alpha^{\beta}(a+bx)e^{-ax-(b/2)x^{2}}}{(1+\alpha^{\beta})\left(1-\frac{e^{-ax-(b/2)x^{2}}}{(1+\alpha^{\beta})}\right)^{2}} dx.
\end{equation*}

Using \eqref{eq5}, we get

\begin{equation*}
\int_{t}^{\infty} x^{r} f(x) dx = \sum_{k=0}^{\infty} \frac{\alpha^{\beta} (1+k)!}{k! (1+\alpha^{\beta})^{k+1}} \int_{q}^{\infty} x^{r} (a+bx) e^{-(k+1)(ax+(b/2)x^{2})} dt,
\end{equation*}

and then using the series expansion

\begin{equation*}
e^{-\frac{(k+1)b}{2}x^{2}} = \sum_{m=0}^{\infty} \frac{(-b)^{m} (k+1)^{m} x^{2m}}{2^{m} m!},
\end{equation*}

we get

\begin{equation}\label{eq8}
\begin{aligned}
\int_{t}^{\infty} x^{r} f(x) dx &= \sum_{k=1}^{\infty} \sum_{m=0}^{\infty} \frac{\alpha^{\beta} k^{m+1} (-b)^{m}}{m! 2^{m} (1+\alpha^{\beta})^{k}} \\
&\quad \times \left\{ \frac{a \Gamma(akt, 2m+r+1)}{(ak)^{2m+r+1}} + \frac{b \Gamma(akt, 2m+r+2)}{(ak)^{2m+r+2}} \right\}.
\end{aligned}
\end{equation}

Therefore

\begin{equation*}
\begin{aligned}
E(X^{r}|X > t) &= \frac{(\alpha^{\beta}+1)e^{ax+(b/2)x^{2}}-1}{\alpha^{\beta}} \sum_{k=1}^{\infty} \sum_{m=0}^{\infty} \frac{\alpha^{\beta} k^{m+1} (-b)^{m}}{m! 2^{m} (1+\alpha^{\beta})^{k}} \\
&\quad \times \left\{ \frac{a \Gamma(akt, 2m+r+1)}{(ak)^{2m+r+1}} + \frac{b \Gamma(akt, 2m+r+2)}{(ak)^{2m+r+2}} \right\}.
\end{aligned}
\end{equation*}

\subsection{Mean Residual Life Function}

The mean residual life function, also known as the expected residual life, is a key reliability and it describes the expected remaining lifetime of a system or component given that it has already survived up to a certain time $t>0$. For a non-negative random variable $X$ (representing lifetime), the mean residual life function of the MLFR distribution is

\begin{equation*}
m(t) = E(X-t|X>t) = \frac{1}{S(x)} \int_{t}^{\infty} x f(x) dx - t,
\end{equation*}

where from \eqref{eq8} with $r=1$

\begin{equation*}
\int_{t}^{\infty} x f(x) dx = \sum_{k=1}^{\infty} \sum_{m=0}^{\infty} \frac{\alpha^{\beta} k^{m+1} (-b)^{m}}{m! 2^{m} (1+\alpha^{\beta})^{k}} \left\{ \frac{a \Gamma(akt, 2m+2)}{(ak)^{2m+2}} + \frac{b \Gamma(akt, 2m+3)}{(ak)^{2m+3}} \right\}.
\end{equation*}

Therefore

\begin{equation*}
m(t) = \frac{1}{S(x)} \left( \sum_{k=1}^{\infty} \sum_{m=0}^{\infty} \frac{\alpha^{\beta} k^{m+1} (-b)^{m}}{m! 2^{m} (1+\alpha^{\beta})^{k}} \left\{ \frac{a \Gamma(akt, 2m+2)}{(ak)^{2m+2}} + \frac{b \Gamma(akt, 2m+3)}{(ak)^{2m+3}} \right\} - t \right).
\end{equation*}

\subsection{Mean Deviations}

The mean absolute deviation of any random variable $X$ from its mean $\mu=\mu_{1}'$ is

\begin{equation*}
\delta_{1} = E(|X-\mu|) = 2\mu F(\mu) - 2\mu + 2\int_{\mu}^{\infty} x f(x) dx.
\end{equation*}

The mean absolute deviation of $X$ from its median $M=Q(1/2)$, is

\begin{equation*}
\delta_{2} = E(|X-M|) = -\mu + 2\int_{M}^{\infty} x f(x) dx,
\end{equation*}

If $X$ has the PDF (\ref{eq:4}), then from \eqref{eq8} with $r=1$, one can get, for $t=\mu$,

\begin{equation*}
\begin{aligned}
\delta_{1} &= 2\mu F(\mu) - 2\mu \\
&\quad + 2\sum_{k=1}^{\infty} \sum_{m=0}^{\infty} \frac{\alpha^{\beta} k^{m+1} (-b)^{m}}{m! 2^{m} (1+\alpha^{\beta})^{k}} \left\{ \frac{a \Gamma(ak\mu, 2m+2)}{(ak)^{2m+2}} + \frac{b \Gamma(ak\mu, 2m+3)}{(ak)^{2m+3}} \right\},
\end{aligned}
\end{equation*}

and for $t=M$,

\begin{equation*}
\delta_{2} = -\mu + 2\sum_{k=1}^{\infty} \sum_{m=0}^{\infty} \frac{\alpha^{\beta} k^{m+1} (-b)^{m}}{m! 2^{m} (1+\alpha^{\beta})^{k}} \left\{ \frac{a \Gamma(akM, 2m+2)}{(ak)^{2m+2}} + \frac{b \Gamma(akM, 2m+3)}{(ak)^{2m+3}} \right\}.
\end{equation*}

\subsection{Bonferroni and Lorenz Curves}

These curves are applied in various discipline scientifics like reliability, medicine and economics. The Bonferroni and Lorenz curves are, respectively, given by

\begin{equation*}
B(p) = \frac{1}{p} - \frac{1}{p\mu} \int_{q}^{\infty} x f(x) dx \quad \text{and} \quad L(p) = 1 - \frac{1}{\mu} \int_{q}^{\infty} x f(x) dx,
\end{equation*}

where $\mu=E(X)$ and $q=Q(p)$. Therefore from \eqref{eq8} with $r=1$, one can get, for $t=q$,

\begin{equation*}
\begin{aligned}
B(p) &= \frac{1}{p} - \frac{1}{p\mu} \sum_{k=1}^{\infty} \sum_{m=0}^{\infty} \frac{\alpha^{\beta} k^{m+1} (-b)^{m}}{m! 2^{m} (1+\alpha^{\beta})^{k}} \\
&\quad \times \left\{ \frac{a \Gamma(akq, 2m+2)}{(ak)^{2m+2}} + \frac{b \Gamma(akq, 2m+3)}{(ak)^{2m+3}} \right\},
\end{aligned}
\end{equation*}

and

\begin{equation*}
\begin{aligned}
L(p) &= 1 - \frac{1}{\mu} \sum_{k=0}^{\infty} \sum_{m=0}^{\infty} \frac{\alpha^{\beta} (k+1)^{m+1} (-b)^{m}}{m! 2^{m} (1+\alpha^{\beta})^{k+1}} \\
&\quad \times \left\{ \frac{a \Gamma(a(k+1)q, 2m+2)}{(a(k+1))^{2m+2}} + \frac{b \Gamma(a(k+1)q, 2m+3)}{(a(k+1))^{2m+3}} \right\}.
\end{aligned}
\end{equation*}

\subsection{Rényi Entropy}

The Rényi entropy, introduced by Alfréd Rényi in 1961 \cite{Renyi1961}, measures the uncertainty. The Rényi entropy is

\begin{equation*}
I_{R}(s) = \frac{1}{1-s} \log \left( \int_{\mathbb{R}} f^{s}(x) dx \right), \quad s>0, s \neq 1.
\end{equation*}

We get the Shannon entropy when $s \to 1$ ( \cite{Shannon1951}). Then, if $X$ has the PDF (\ref{eq:4}), we have

\begin{equation*}
I_{R}(s) = \frac{1}{1-s} \log \left( \int_{0}^{\infty} f^{s}(x) dx \right), \quad s>0, s \neq 1,
\end{equation*}

where

\begin{equation*}
f^{s}(x) = \frac{\alpha^{s\beta} (a+bx)^{s} e^{-sax-\frac{sb}{2}x^{2}}}{(1+\alpha^{\beta})^{s} \left( 1 - \frac{e^{-ax-(b/2)x^{2}}}{(1+\alpha^{\beta})} \right)^{2s}}.
\end{equation*}

By applying \eqref{eq5}, we get

\begin{equation*}
f^{s}(x) = \frac{\alpha^{s\beta}}{(2s-1)} \sum_{k=0}^{\infty} \frac{(2s-1+k)!}{k! (1+\alpha^{\beta})^{s+k}} (a+bx)^{s} e^{-(s+k)(ax+\frac{b}{2}x^{2})},
\end{equation*}

and then using the following series expansion

\begin{equation*}
e^{-\frac{(s+k)b}{2}x^{2}} = \sum_{m=0}^{\infty} \frac{(-b)^{m} (k+s)^{m} x^{2m}}{2^{m} m!},
\end{equation*}

we get

\begin{equation*}
\begin{aligned}
I_{R}(s) &= \frac{1}{1-s} \log \Bigg( \frac{\alpha^{s\beta}}{(2s-1)} \sum_{k=0}^{\infty} \sum_{m=0}^{\infty} \frac{(2s-1+k)! (-b)^{m} (s+k)^{m}}{k! (1+\alpha^{\beta})^{s+k} m! 2^{m}} \\
&\quad \times \int_{0}^{\infty} (a+bx)^{s} x^{2m} e^{-a(s+k)x} dx \Bigg).
\end{aligned}
\end{equation*}

Substituting $t=a+bx$ in the last integral and then by the binomial expansion of $(a+bx)^{s}$, we get

\begin{equation*}
\int_{0}^{\infty} (a+bx)^{s} x^{2m} e^{-a(s+k)x} dx = \frac{e^{\frac{a^{2}(s+k)}{b}}}{b^{2m+1}} \sum_{l=0}^{2m} (-a)^{l} \int_{a}^{\infty} t^{2m-l+s} e^{-\frac{a(s+k)t}{b}} dt.
\end{equation*}

Finally, by making the substitution $z=\frac{a(s+k)}{b}t$ into the previous integral, we get

\begin{equation*}
\begin{aligned}
I_{R}(s) &= \frac{1}{1-s} \log \Bigg\{ \frac{\alpha^{s\beta}}{(2s-1)} \sum_{k=0}^{\infty} \sum_{m=0}^{\infty} \sum_{l=0}^{2m} \frac{(-b)^{m} (2s-1+k)! (s+k)^{m} (-1)^{l}}{k! (1+\alpha^{\beta})^{s+k} m! 2^{m} [a(s+k)]^{2m+s-l+1}} \\
&\quad \times a^{l} b^{s-l} e^{\frac{a^{2}(s+k)}{b}} \Gamma\left(2m+s-l+1, \frac{a^{2}(s+k)}{b}\right) \Bigg\}.
\end{aligned}
\end{equation*}

where $\Gamma(\cdot,\cdot)$ is the upper incomplete gamma function.

\subsection{Reliability}

In the stress-strength model, reliability is defined as the probability that a system's strength exceeds the stress applied to it, i.e., $R=P(X_{2}<X_{1})$, where $X_{1}$ and $X_{2}$ are independent, with $X_{1}$ being the strength of the system and $X_{2}$ the stress applied to the system. So, reliability $R$ represents the probability that the system will not fail, i.e., the strength $X_{1}$ is greater than the stress $X_{2}$. Suppose $X_{1}$ and $X_{2}$ have the different MLFR model parameters, then

\begin{equation*}
\begin{aligned}
R &= \int_{0}^{\infty} f(x;\alpha_{1},\beta_{1},a_{1},b_{1}) F(x;\alpha_{2},\beta_{2},a_{2},b_{2}) dx \\
&= \frac{\alpha_{1}^{\beta_{1}}}{(1+\alpha_{1}^{\beta_{1}})} \int_{0}^{\infty} \frac{(a_{1}+b_{1}x)(e^{-a_{1}x-\frac{b_{1}}{2}x^{2}} - e^{-a_{1}x-\frac{b_{1}}{2}x^{2}} e^{-a_{2}x-\frac{b_{2}}{2}x^{2}})}{\left(1-\frac{e^{-a_{1}x-\frac{b_{1}}{2}x^{2}}}{1+\alpha_{1}^{\beta_{1}}}\right)^{2} \left(1-\frac{e^{-a_{2}x-\frac{b_{2}}{2}x^{2}}}{1+\alpha_{2}^{\beta_{2}}}\right)} dx.
\end{aligned}
\end{equation*}

Using \eqref{eq5}, and after some algebraic manipulation, we obtain

\begin{equation*}
\begin{aligned}
R &= \sum_{k=0}^{\infty} \sum_{\ell=0}^{\infty} \frac{\alpha_{1}^{\beta_{1}}(k+1)}{(1+\alpha_{1}^{\beta_{1}})^{k+1} (1+\alpha_{2}^{\beta_{2}})^{\ell}} \Bigg\{ a_{1} \int_{0}^{\infty} e^{-\ell(a_{2}x+\frac{b_{2}}{2}x^{2})} e^{-(k+1)(a_{1}x+\frac{b_{1}}{2}x^{2})} dx \\
&\quad - b_{1} \int_{0}^{\infty} x e^{-(\ell+1)(a_{2}x+\frac{b_{2}}{2}x^{2})} e^{-(k+1)(a_{1}x+\frac{b_{1}}{2}x^{2})} dx \Bigg\}.
\end{aligned}
\end{equation*}

From \eqref{eq6}, we get

\begin{equation*}
\begin{aligned}
R &= \sum_{k=1}^{\infty} \sum_{\ell=1}^{\infty} \sum_{m=1}^{\infty} \frac{(-1)^{m-1} \alpha_{1}^{\beta_{1}} k}{2(m-1)! (1+\alpha_{1}^{\beta_{1}})^{k-1} (1+\alpha_{2}^{\beta_{2}})^{\ell-1}} \\
&\quad \times \Bigg\{ \frac{a_{1} (ka_{1}+(\ell-1)a_{2})^{m-1} \Gamma\left(\frac{m}{2}\right)}{\left(\frac{kb_{1}+(\ell-1)b_{2}}{2}\right)^{(m/2)}} + \frac{b_{1} (\ell a_{1}+ma_{2})^{m-1} \Gamma\left(\frac{m+1}{2}\right)}{\left(\frac{b_{1}\ell+mb_{2}}{2}\right)^{((m+1)/2)}} \Bigg\}.
\end{aligned}
\end{equation*}

\subsection{Order Statistics}

Order statistics, denoted by $X_{1,n},\ldots,X_{n,n}$ represent the sorted values of a random sample $X_{1},\ldots,X_{n}$. Their distribution plays a critical role in reliability, especially as the minimum and maximum can describe lifetimes of series and parallel systems, respectively. According to \cite{ArnoldBalakrishnanNagaraja2008}, the CDF of $k$th order statistic $X_{k,n}$ is

\begin{equation*}
F_{k}(x) = \sum_{j=k}^{n} \sum_{\ell=0}^{n-j} (-1)^{\ell} \binom{n}{j} \binom{n-j}{\ell} (F(x))^{j+\ell},
\end{equation*}

and the PDF of $X_{k,n}$ is

\begin{equation*}
f_{k}(x) = \frac{n!}{(n-k)!(k-1)!} \sum_{\ell=0}^{n-k} (-1)^{\ell} \binom{n-k}{l} f(x) (F(x))^{k+\ell-1}, \quad k=1,\ldots,n.
\end{equation*}

If $X_{1}$ has the MLFR model, therefore using \eqref{eq:3} and \eqref{eq:4}, we get

\begin{equation*}
F_{k}(x) = \sum_{j=k}^{n} \sum_{\ell=0}^{n-j} (-1)^{l} \binom{n}{j} \binom{n-j}{l} \frac{(1+\alpha^{\beta})^{j+l} (1-e^{-ax-(b/2)x^{2}})^{j+l}}{(\alpha^{\beta}+1-e^{-ax-(b/2)x^{2}})^{j+l}},
\end{equation*}

and

\begin{equation*}
f_{k}(x) = \frac{n! \alpha^{\beta} (a+bx) e^{-ax-(b/2)x^{2}}}{(n-k)!(k-1)!} \sum_{\ell=0}^{n-k} (-1)^{\ell} \binom{n-k}{\ell} \frac{(1+\alpha^{\beta})^{k+\ell} (1-e^{-ax-(b/2)x^{2}})^{k+\ell-1}}{(1+\alpha^{\beta}-e^{-ax-(b/2)x^{2}})^{k+\ell+1}}.
\end{equation*}

\section{Parameter Estimation}\label{sec4}

The maximum likelihood approach was employed to estimate the unknown parameters of the MLFR distribution. The log-likelihood function for the parameters $\alpha,\beta,a,b$ based on the observed values $x_{1},x_{2},\ldots,x_{n}$ of $X_{1},X_{2},\ldots,X_{n}$ where the $X_{i}$'s are independent and identically distributed random variables having the PDF \eqref{eq:4}, is given by:

\begin{equation*}
\begin{aligned}
L(\alpha,\beta,a,b) &= n\beta \log(\alpha) + n \log(1+\alpha^{\beta}) + \sum_{i=1}^{n} \log(a+bx_{i}) - a \sum_{i=1}^{n} \log x_{i} - \frac{b}{2} \sum_{i=1}^{n} x_{i}^{2} \\
&\quad - 2 \sum_{i=1}^{n} \log(1+\alpha^{\beta} - e^{-ax-(b/2)x^{2}}).
\end{aligned}
\end{equation*}

The first partial derivatives of $L$, with respect to $\alpha,\beta,a$ and $b$ are

\begin{equation*}
\begin{aligned}
\frac{\partial L(\alpha,\beta,a,b)}{\partial \alpha} &= \frac{n\beta(2\alpha^{\beta}+1)}{\alpha(\alpha^{\beta}+1)} - 2\sum_{i=1}^{n} \frac{\beta \alpha^{\beta-1}}{1+\alpha^{\beta} - e^{-ax_{i}-(b/2)x_{i}^{2}}}, \\
\frac{\partial L(\alpha,\beta,a,b)}{\partial \beta} &= \frac{n(2\alpha^{\beta}+1)\log(\alpha)}{\alpha^{\beta}+1} - 2\sum_{i=1}^{n} \frac{\alpha^{\beta} \log(\alpha)}{1+\alpha^{\beta} - e^{-ax_{i}-(b/2)x_{i}^{2}}}, \\
\frac{\partial L(\alpha,\beta,a,b)}{\partial a} &= \sum_{i=1}^{n} \frac{1}{a+bx_{i}} + 2\sum_{i=1}^{n} \frac{x_{i}}{(1+\alpha^{\beta})e^{-ax_{i}-(b/2)x_{i}^{2}} - 1},
\end{aligned}
\end{equation*}

and

\begin{equation*}
\frac{\partial L(\alpha,\beta,a,b)}{\partial b} = \sum_{i=1}^{n} \frac{x_{i}}{a+bx_{i}} - \frac{1}{2} \sum_{i=1}^{n} x_{i}^{2} - \sum_{i=1}^{n} \frac{x_{i}^{2}}{(1+\alpha^{\beta})e^{-ax_{i}-(b/2)x_{i}^{2}} - 1}.
\end{equation*}

These equations cannot be solved analytically, due to the nonlinear and complex structure. Therefore, numerical optimization methods, such as the Newton-Raphson algorithm, must be employed to solve the system of equations and obtain the maximum likelihood estimates (MLEs) of $\alpha,\beta,a$ and $b$.

It is well known that, the MLEs of $\alpha,\beta,a$ and $b$, are jointly asymptotically normal with mean equal 0 and covariance-matrix $I^{-1}(\alpha,\beta,a,b)$, where

\begin{equation*}
I(\alpha,\beta,a,b) = - \begin{pmatrix}
\frac{\partial^{2} L}{\partial \alpha^{2}} & \frac{\partial^{2} L}{\partial \alpha \partial \beta} & \frac{\partial^{2} L}{\partial \alpha \partial a} & \frac{\partial^{2} L}{\partial \alpha \partial b} \\
\frac{\partial^{2} L}{\partial \alpha \partial \beta} & \frac{\partial^{2} L}{\partial \beta^{2}} & \frac{\partial^{2} L}{\partial \beta \partial a} & \frac{\partial^{2} L}{\partial \beta \partial b} \\
\frac{\partial^{2} L}{\partial \alpha \partial a} & \frac{\partial^{2} L}{\partial \beta \partial a} & \frac{\partial^{2} L}{\partial a^{2}} & \frac{\partial^{2} L}{\partial a \partial b} \\
\frac{\partial^{2} L}{\partial \alpha \partial b} & \frac{\partial^{2} L}{\partial \beta \partial b} & \frac{\partial^{2} L}{\partial a \partial b} & \frac{\partial^{2} L}{\partial b^{2}}
\end{pmatrix}.
\end{equation*}

The analytical expressions for the components of $I(\alpha,\beta,a,b)$ are available from the author request. This result, allows us to construct the approximate confidence intervals of $\alpha,\beta,a$ and $b$ which are given by

\begin{equation*}
\alpha \pm Z_{(\zeta/2)} \sqrt{V(\alpha)}, \quad \beta \pm Z_{(\zeta/2)} \sqrt{V(\beta)}, \quad a \pm Z_{(\zeta/2)} \sqrt{V(a)} \quad \text{and} \quad b \pm Z_{(\zeta/2)} \sqrt{V(b)},
\end{equation*}

where $V(\cdot)$ represents the diagonal component of $I^{-1}(\alpha,\beta,a,b)$ whereas $Z_{\zeta/2}$ represents the $100(1-\zeta/2)$-th percentile of the standard normal distribution.

\section{Simulation study}\label{sec5}

In order to evaluate the performance of the MLEs of the MLFR distribution, a simulation study is conducted by means of the statistical software R. We use equation (\ref{eq7})  to generate the random samples from the MLFR distribution. We repeat the simulation 1000 times for each combination of sample sizes $n=20,50,100,200,500$, and $1000$. The following scenarios of true parameters $\alpha$, $\beta$, $a$ and $b$ are considered:

\begin{itemize}
    \item Scenario I: $\alpha=1.5$, $\beta=0.1$, $a=0.75$, $b=0.25$,
    \item Scenario II: $\alpha=0.25$, $\beta=0.5$, $a=0.8$, $b=0.75$,
    \item Scenario III: $\alpha=3$, $\beta=0.25$, $a=1.2$, $b=1$.
\end{itemize}

The performance of the MLEs is assessed using the bias and the mean squared errors (MSE), which are given by:

\[
\text{Bias} = \frac{1}{1000} \sum_{i=1}^{1000} (\hat{\lambda}_{i} - \lambda) \quad \text{and} \quad \text{MSE} = \frac{1}{1000} \sum_{i=1}^{1000} (\hat{\lambda}_{i} - \lambda)^2,
\]

where $\lambda = \{\alpha, \beta, a, b\}$. According to Table \ref{tab1}, as the sample size $n$ increases, both the bias and MSE of the MLEs of the MLFR distribution converge to zero. This indicates that the MLEs perform well in finite samples and exhibit desirable large-sample properties, such as asymptotic unbiasedness and consistency.

\begin{table}[ht!]
\centering
\caption{Bias and MSE of the MLEs.}
\label{tab1}
\begin{tabular}{ccccccccc}
\hline
& & \multicolumn{2}{c}{Scenario I} & \multicolumn{2}{c}{Scenario II} & \multicolumn{2}{c}{Scenario III} \\ \cline{3-8}
Sample size & Parameter & Bias & MSE & Bias & MSE & Bias & MSE \\ \hline
\multirow{4}{*}{$n=20$} & $\alpha$ & 0.2859 & 1.2857 & 0.1046 & 0.1307 & 1.1307 & 2.3541 \\
                         & $\beta$  & 0.1130 & 0.1346 & 0.4587 & 0.4671 & 0.1976 & 0.3750 \\
                         & $a$      & 0.0399 & 0.2840 & 0.0439 & 0.4603 & 0.0942 & 0.4929 \\
                         & $b$      & 0.0953 & 0.3211 & 0.3687 & 0.4497 & 0.4066 & 0.7514 \\ \hline
\multirow{4}{*}{$n=50$} & $\alpha$ & 0.1308 & 0.9072 & 0.0616 & 0.1179 & 0.5855 & 2.0534 \\
                         & $\beta$  & 0.1058 & 0.1131 & 0.4410 & 0.4601 & 0.1216 & 0.2844 \\
                         & $a$      & 0.0177 & 0.1627 & 0.0379 & 0.4056 & 0.0618 & 0.3536 \\
                         & $b$      & 0.0430 & 0.2244 & 0.2346 & 0.3993 & 0.2122 & 0.7176 \\ \hline
\multirow{4}{*}{$n=100$} & $\alpha$ & 0.0495 & 0.4751 & 0.0423 & 0.0983 & 0.0756 & 2.0036 \\
                         & $\beta$  & 0.1014 & 0.1047 & 0.4215 & 0.4460 & 0.1120 & 0.2727 \\
                         & $a$      & 0.0047 & 0.1113 & 0.0351 & 0.3371 & 0.0259 & 0.2584 \\
                         & $b$      & 0.0176 & 0.1222 & 0.1614 & 0.3394 & 0.0255 & 0.7013 \\ \hline
\multirow{4}{*}{$n=200$} & $\alpha$ & 0.0056 & 0.3397 & 0.0156 & 0.0794 & 0.0551 & 1.7264 \\
                         & $\beta$  & 0.1012 & 0.1027 & 0.3989 & 0.4129 & 0.1019 & 0.2675 \\
                         & $a$      & 0.0031 & 0.0734 & 0.0254 & 0.3288 & 0.0143 & 0.2015 \\
                         & $b$      & 0.0044 & 0.0856 & 0.0700 & 0.2696 & 0.0203 & 0.6148 \\ \hline
\multirow{4}{*}{$n=500$} & $\alpha$ & 0.0021 & 0.2022 & 0.0051 & 0.0649 & 0.0119 & 1.2447 \\
                         & $\beta$  & 0.1005 & 0.1012 & 0.0948 & 0.1036 & 0.0948 & 0.2341 \\
                         & $a$      & 0.0023 & 0.0503 & 0.0031 & 0.3058 & 0.0086 & 0.1263 \\
                         & $b$      & 0.0007 & 0.0507 & 0.0064 & 0.2213 & 0.0144 & 0.4430 \\ \hline
\multirow{4}{*}{$n=1000$} & $\alpha$ & 0.0012 & 0.1253 & 0.0013 & 0.0402 & 0.0098 & 1.0017 \\
                         & $\beta$  & 0.0857 & 0.0971 & 0.0697 & 0.0922 & 0.0686 & 0.1650 \\
                         & $a$      & 0.0010 & 0.0331 & 0.0012 & 0.2104 & 0.0061 & 0.1002 \\
                         & $b$      & 0.0004 & 0.0309 & 0.0045 & 0.1569 & 0.0098 & 0.2829 \\ \hline
\end{tabular}
\end{table}
\section{Survival data analysis}\label{sec6}

Two famous real survival time data are analyzed in this section to examine the flexibility and competency of our proposed distribution. For these data sets, the fit of the introduced MFLR distribution is compared with the fit of some new models developed recently by the Modi family. The competitive models are:

\begin{itemize}
    \item Modi Rayleigh (MR) distribution (\cite{KumarMeenaShukla2025}) with PDF
    \[
    f(x)=\frac{\alpha^{\beta}(1+\alpha^{\beta})\frac{x}{\sigma^2}e^{-\frac{x^2}{2\sigma^2}}}{\left(1+\alpha^{\beta}-e^{-\frac{x^2}{2\sigma^2}}\right)^2}, \quad x, \alpha, \beta, \sigma>0.
    \]
    
    \item Modi Weibull (MW) distribution (\cite{KumawatModiNagar2023}) with PDF
    \[
    f(x)=\frac{a\alpha^{\beta}(1+\alpha^{\beta})x^{a-1}e^{-(x/b)^a}}{b^a\left(\alpha^{\beta}+1-e^{-(x/b)^a}\right)^2}, \quad x, \alpha, \beta, a, b>0.
    \]
    
    \item Modi exponential (ME) distribution (\cite{ModiKumarSingh2020}) with PDF
    \[
    f(x)=\frac{a\alpha^{\beta}(1+\alpha^{\beta})e^{-ax}}{\left(\alpha^{\beta}+1-e^{-ax}\right)^2}, \quad x, \alpha, \beta, a>0.
    \]
    
    \item Modi Fréchet (MF) distribution (\cite{AkhilaBabu2025}) with PDF
    \[
    f(x)=\frac{ab^{a}\alpha^{\beta}x^{-a-1}(1+\alpha^{\beta})e^{-(b/x)^a}}{\left(\alpha^{\beta}+e^{-(b/x)^a}\right)^2}, \quad x, \alpha, \beta, a, b>0.
    \]
\end{itemize}

The comparison is based on several well-known model selection criteria: minus twice the maximized log-likelihood ($-2\log L$), Akaike information criterion (AIC), Bayesian information criterion (BIC), Consistent Akaike information criterion (CAIC), and Hannan-Quinn information criterion (HQC). Additionally, we employ the Kolmogorov-Smirnov (K-S), Cramer-von-Mises (CvM) and Anderson-Darling (AD) statistics with their $p$-values. We select the best model, which has the largest $p$-values and the smallest values of $-2\log L$, AIC, BIC, CAIC, HQIC, K-S, CvM and AD statistics.

\subsection*{First data set: bladder cancer remission times}

This real data set is taken from \cite{LeeWang2003} and refers to the survival times, in months, of 128 individuals diagnosed with bladder cancer. The data set is: 3.88, 7.39, 10.34, 14.83, 34.26, 0.90, 2.69, 4.18, 5.34, 7.59, 10.66, 15.96, 36.66, 1.05, 2.69, 4.23, 5.41, 7.62, 10.75, 16.62, 43.01, 1.19, 2.75, 4.26, 5.41, 7.63, 17.12, 46.12, 1.26, 2.83, 4.33, 5.49, 7.66, 11.25, 17.14, 79.05, 5.32, 1.35, 2.87, 5.62, 7.87, 11.64, 17.36, 1.40, 3.02, 4.34, 5.71, 7.93, 0.08, 2.09, 3.48, 4.87, 6.94, 8.66, 13.11, 23.63, 0.20, 2.23, 3.5, 4.98, 6.97, 9.02, 13.29, 0.40, 2.26, 3.57, 5.06, 7.09, 9.22, 13.80, 25.74, 0.50, 2.46, 3.64, 5.09, 7.26, 9.47, 14.24, 25.82, 0.51, 2.54, 3.70, 5.17, 7.28, 9.74, 14.76, 26.31, 0.81, 2.62, 3.82, 5.32, 7.32, 10.06, 14.77, 32.15, 2.64, 11.79, 18.10, 1.46, 4.40, 5.85, 8.26, 11.98, 19.13, 1.76, 3.25, 4.50, 6.25, 8.37, 12.02, 2.02, 3.31, 4.51, 6.54, 8.53, 12.03, 20.28, 2.02, 3.36, 6.76, 12.07, 21.73, 2.00, 3.36, 6.93, 8.65, 12.63 and 22.69. This data set has been studied by several authors like \cite{Benkhelifa2017} and \cite{KumawatModiNagar2023}.

\begin{figure}[ht!]
\centering
\includegraphics[scale=0.6]{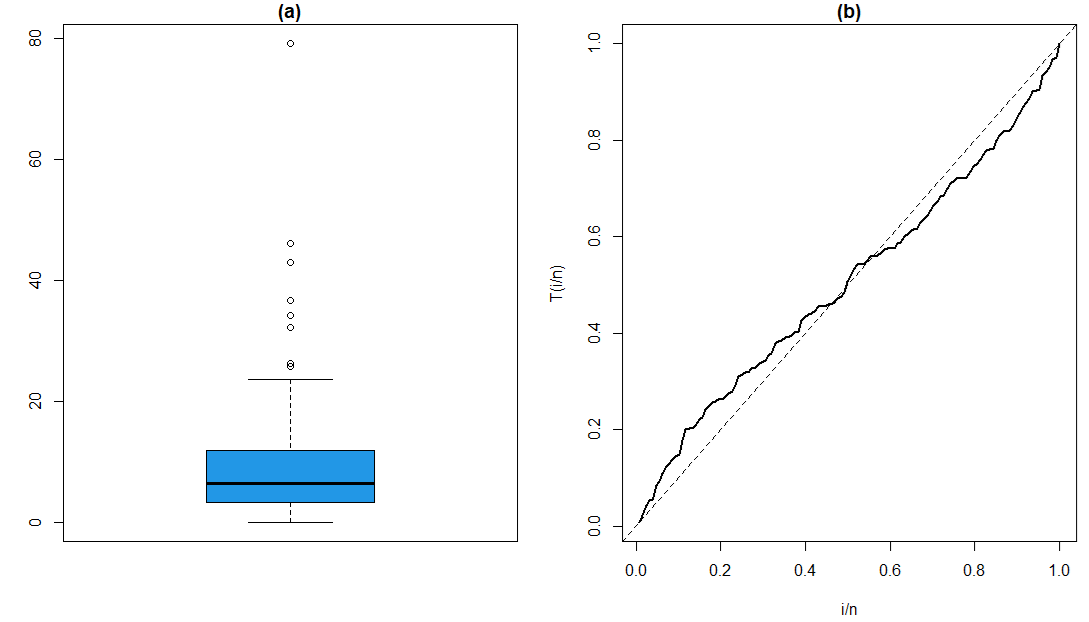}
\caption{(a) Box plot and (b) TTT plot for first data.}
\label{fig3}
\end{figure}

Some descriptive statistics are displayed in Table \ref{tab2}. As shown in this table, the data set exhibits a skewness of 3.286 and a kurtosis of 18.481. The high positive skewness indicates that the distribution is strongly skewed to the right, with a long tail extending toward higher values. The kurtosis value, significantly greater than 3, suggests that the data are leptokurtic. Additionally, the distribution is described as unimodal, indicating a single prominent peak in the data. Figure \ref{fig3}(a) presents the boxplot of the data, which reveals the presence of more than six outliers. These extreme values are consistent with the observed high skewness and kurtosis, reinforcing the departure from symmetry and normality. Figure \ref{fig3}(b) displays the TTT (total time on test) plot, which takes a convex shape followed by a concave shape. This corresponds to an upside-down bathtub hazard rate. Hence, the MLFR distribution is suitable to model this lifetime data.

\begin{figure}[ht!]
\centering
\includegraphics[scale=0.4]{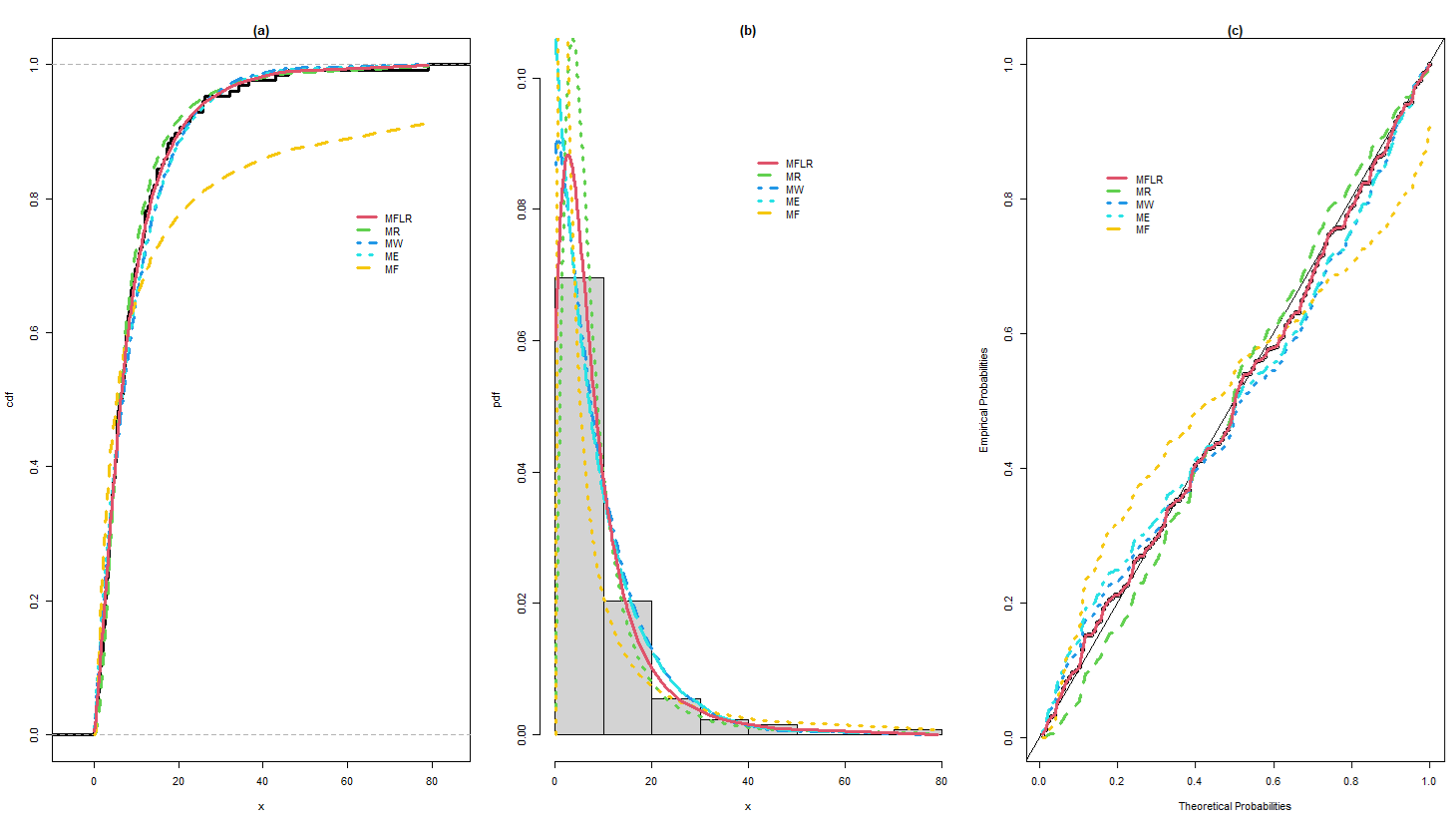}
\caption{ (a) ECDF with the fitted CDFs, (b) Histogram with the fitted PDFs and (c) PP plot for first data.}
\label{fig4}
\end{figure}

Using the R software, the MLEs for all candidate models were obtained by employing the mle2 function from the bbmle package. Table \ref{tab3} presents the MLEs along with the corresponding values of $-2\log L$, AIC, BIC, CAIC and HQIC whereas the Table \ref{tab4} gives the K-S, CvM and AD statistics, together with their respective $p$-values. The results in these tables demonstrate that the MLFR model provides the best fit among all competing models, because it has the lowest values of the information criteria and the highest $p$-values of the goodness-of-fit tests. On the other hand, Figure \ref{fig4} displays the fitted CDF, PDF, and PP plots for the MLFR distribution. These graphical representations show that the MLFR model closely follows the empirical CDF, the histogram of the data, and the diagonal line in the PP plot, further confirming its superior performance and adequacy in modeling the given data set.

\begin{table}[ht!]
\centering
\caption{Descriptive statistics for the first data.}
\label{tab2}
\begin{tabular}{|c|c|c|c|c|c|c|c|c|c|}
\hline
Min & Q1 & Q2 & Mean & Q3 & Max & Std.Dev & Skewness & Kurtosis \\
\hline
0.080 & 3.348 & 6.395 & 9.365 & 11.838 & 79.050 & 10.508 & 3.286 & 18.481 \\
\hline
\end{tabular}
\end{table}

\begin{table}[ht!]
\centering
\caption{MLEs, $-2\log L$, AIC, BIC, CAIC, HQIC for the first data.}
\label{tab3}
\begin{tabular}{|l|l|c|c|c|c|c|}
\hline
Model & MLE(s) & $-2\log L$ & AIC & BIC & CAIC & HQIC \\
\hline
\multirow{4}{*}{MFLR} & $\alpha=0.0145$ & \multirow{4}{*}{818.2356} & \multirow{4}{*}{826.2356} & \multirow{4}{*}{837.6437} & \multirow{4}{*}{841.6437} & \multirow{4}{*}{830.8708} \\
 & $\beta=0.6912$ & & & & & \\
 & $a=0.0029$ & & & & & \\
 & $b=0.0015$ & & & & & \\
\hline
\multirow{3}{*}{MR} & $\alpha=0.2637$ & \multirow{3}{*}{826.5442} & \multirow{3}{*}{832.5442} & \multirow{3}{*}{841.1003} & \multirow{3}{*}{844.1003} & \multirow{3}{*}{836.0206} \\
 & $\beta=3.4104$ & & & & & \\
 & $\sigma=42.4147$ & & & & & \\
\hline
\multirow{4}{*}{MW} & $\alpha=4.0945$ & \multirow{4}{*}{828.1591} & \multirow{4}{*}{836.1591} & \multirow{4}{*}{847.5673} & \multirow{4}{*}{851.5673} & \multirow{4}{*}{840.7943} \\
 & $\beta=15.9590$ & & & & & \\
 & $a=1.0476$ & & & & & \\
 & $b=9.5591$ & & & & & \\
\hline
\multirow{3}{*}{ME} & $\alpha=4.4228$ & \multirow{3}{*}{828.6653} & \multirow{3}{*}{834.6653} & \multirow{3}{*}{843.2214} & \multirow{3}{*}{846.2214} & \multirow{3}{*}{838.1417} \\
 & $\beta=4.9330$ & & & & & \\
 & $a=0.1067$ & & & & & \\
\hline
\multirow{4}{*}{MF} & $\alpha=20.3620$ & \multirow{4}{*}{887.9469} & \multirow{4}{*}{895.9469} & \multirow{4}{*}{907.355} & \multirow{4}{*}{911.355} & \multirow{4}{*}{900.5793} \\
 & $\beta=2.3798$ & & & & & \\
 & $a=0.7519$ & & & & & \\
 & $b=3.2588$ & & & & & \\
\hline
\end{tabular}
\end{table}

\begin{table}[ht!]
\centering
\centering
\caption{K-S, CvM and AD statistics with their $p$-values for the first data.}
\label{tab4}
\begin{tabular}{|l|c|c|c|c|c|c|}
\hline
Model & K-S & $p$-value & CvM & $p$-value & AD & $p$-value \\
\hline
MFLR & 0.0417 & 0.9792 & 0.0249 & 0.9901 & 0.1824 & 0.9945 \\
MR & 0.0508 & 0.8959 & 0.0821 & 0.6806 & 0.8531 & 0.4441 \\
MW & 0.0699 & 0.5583 & 0.1530 & 0.3811 & 0.9533 & 0.3826 \\
ME & 0.0830 & 0.3415 & 0.1779 & 0.3149 & 1.1684 & 0.2798 \\
MF & 0.1407 & 0.0126 & 0.9772 & 0.0027 & 6.1087 & 0.0008 \\
\hline
\end{tabular}
\end{table}

\subsection*{Second data set: Infected guinea pigs data}

\begin{figure}[ht!]
\centering
\includegraphics[scale=0.7]{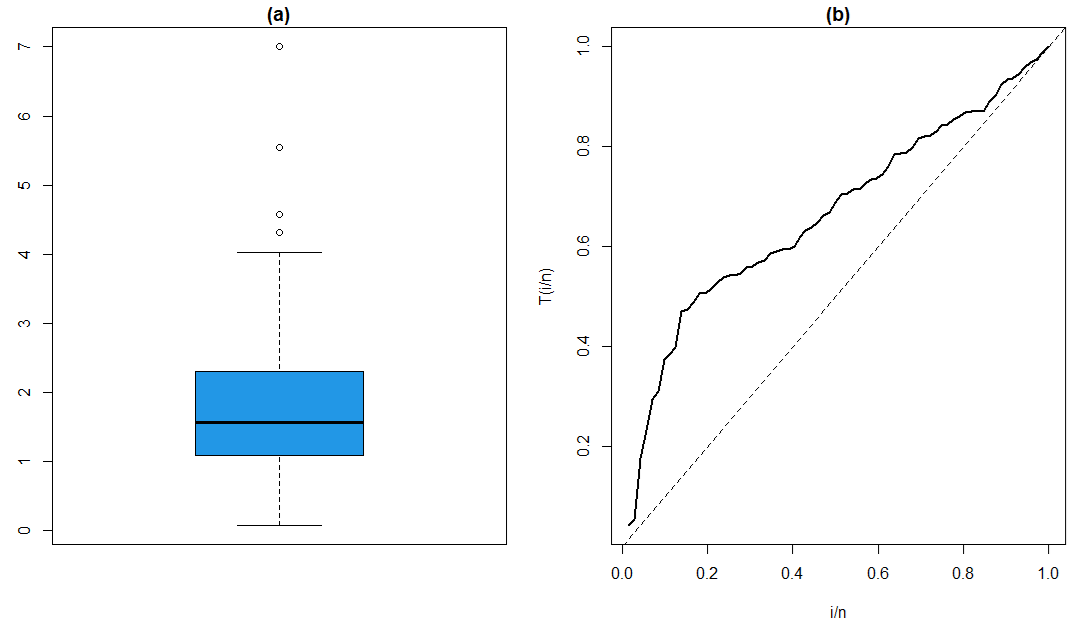}
\caption{(a) Box plot and (b) TTT plot for second data.}
\label{fig5}
\end{figure}

Gross and Clark \cite{GrossClark1975} were the first to analyze this data, which records the survival times (in days) of 72 guinea pigs after being infected with virulent tubercle bacilli. The data are: 2.54, 1.08, 0.1, 0.56, 0.72, 0.44, 0.59, 0.33, 0.74, 0.77, 0.92, 0.93, 0.96, 1, 1, 1.02, 1.05, 1.07, 7, 0.08, 1.08, 1.09, 1.12, 1.13, 1.15, 1.16, 1.2, 1.21, 1.22, 1.22, 1.24, 1.3, 1.34, 1.36, 1.39, 1.44, 1.46, 1.53, 1.59, 1.6, 1.63, 1.63, 1.68, 1.71, 1.72, 1.76, 1.83, 1.95, 1.96, 1.97, 2.02, 2.13, 2.15, 2.16, 2.22, 2.3, 2.31, 2.4, 2.45, 2.51, 2.53, 2.54, 2.78, 2.93, 3.27, 3.42, 3.47, 3.61, 4.02, 4.32, 4.58, 5.55. Some descriptive statistics are presented in Table \ref{tab5}. Based on this table, we observe that the data set is unimodal and positively skewed (indicating a right-skewed distribution). The concave shape of the TTT plot in Figure \ref{fig5}(b) suggests an increasing hazard rate, implying that the introduced distribution is suitable to model this data.

\begin{figure}[ht!]
\centering
\includegraphics[scale=0.4]{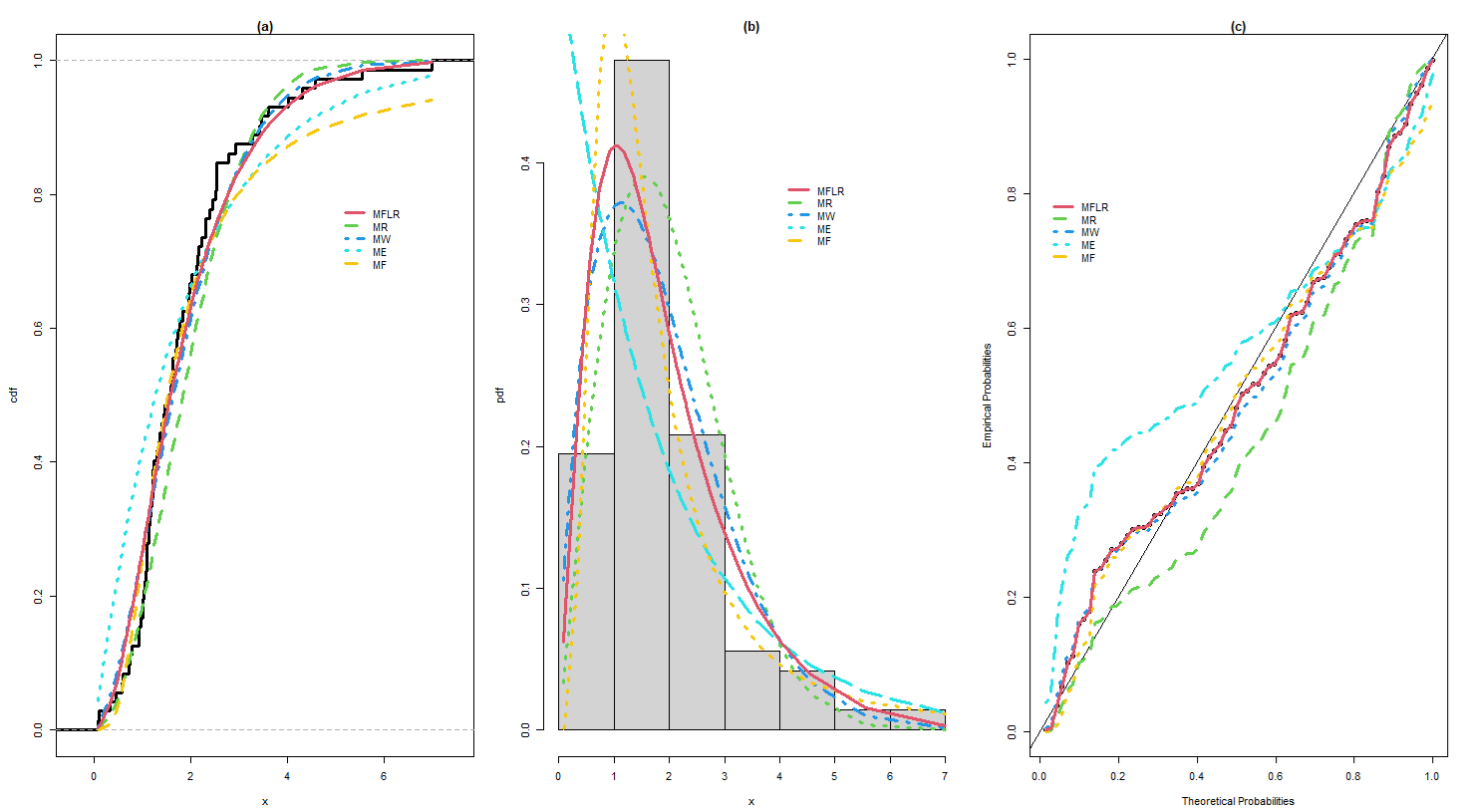}
\caption{ (a) ECDF with the fitted CDFs, (b) Histogram with the fitted PDFs and (c) PP plot for second data.}
\label{fig6}
\end{figure}

Table \ref{tab6} presents the MLEs along with the values of $-2\log L$, AIC, BIC, CAIC and HQIC. The Table \ref{tab7} gives the K-S, CvM and AD statistics with their $p$-values. Based on these results, the MLFR distribution provides the best fit among all the competing models for second data set. This conclusion is further supported by Figure \ref{fig6}, which illustrate that the proposed model offers a superior fit to the data.

\begin{table}[ht!]
\centering
\caption{Descriptive statistics for the second data.}
\label{tab5}
\begin{tabular}{|c|c|c|c|c|c|c|c|c|c|}
\hline
Min & Q1 & Q2 & Mean & Q3 & Max & Std.Dev & Skewness & Kurtosis \\
\hline
0.080 & 1.080 & 1.560 & 1.837 & 2.303 & 7.000 & 1.216 & 1.755 & 7.152 \\
\hline
\end{tabular}
\end{table}

\begin{table}[ht!]
\centering
\caption{MLEs, $-2\log L$, AIC, BIC, CAIC, HQIC for the second data.}
\label{tab6}
\begin{tabular}{|l|l|c|c|c|c|c|}
\hline
Model & MLE(s) & $-2\log L$ & AIC & BIC & CAIC & HQIC \\
\hline
\multirow{4}{*}{MFLR} & $\alpha=0.2489$ & \multirow{4}{*}{205.5156} & \multirow{4}{*}{213.5156} & \multirow{4}{*}{222.6222} & \multirow{4}{*}{226.6222} & \multirow{4}{*}{217.141} \\
 & $\beta=0.6584$ & & & & & \\
 & $a=0.0019$ & & & & & \\
 & $b=0.1983$ & & & & & \\
\hline
\multirow{3}{*}{MR} & $\alpha=231.180$ & \multirow{3}{*}{214.9544} & \multirow{3}{*}{220.9544} & \multirow{3}{*}{227.7844} & \multirow{3}{*}{230.7844} & \multirow{3}{*}{223.6734} \\
 & $\beta=19.4690$ & & & & & \\
 & $\sigma=1.5541$ & & & & & \\
\hline
\multirow{4}{*}{MW} & $\alpha=288.14$ & \multirow{4}{*}{208.0336} & \multirow{4}{*}{216.0336} & \multirow{4}{*}{225.1403} & \multirow{4}{*}{229.1403} & \multirow{4}{*}{219.659} \\
 & $\beta=3.3678$ & & & & & \\
 & $a=1.6173$ & & & & & \\
 & $b=2.0559$ & & & & & \\
\hline
\multirow{3}{*}{ME} & $\alpha=4.5832$ & \multirow{3}{*}{231.5558} & \multirow{3}{*}{237.5558} & \multirow{3}{*}{244.3858} & \multirow{3}{*}{247.3858} & \multirow{3}{*}{240.2748} \\
 & $\beta=4.9496$ & & & & & \\
 & $a=0.5440$ & & & & & \\
\hline
\multirow{4}{*}{MF} & $\alpha=0.6104$ & \multirow{4}{*}{225.4629} & \multirow{4}{*}{233.4629} & \multirow{4}{*}{242.5695} & \multirow{4}{*}{246.5695} & \multirow{4}{*}{236.4363} \\
 & $\beta=12.469$ & & & & & \\
 & $a=0.3474$ & & & & & \\
 & $b=287.25$ & & & & & \\
\hline
\end{tabular}
\end{table}

\begin{table}[ht!]
\centering
\caption{K-S, CvM and AD statistics with their $p$-values for the second data.}
\label{tab7}
\begin{tabular}{|l|c|c|c|c|c|c|}
\hline
Model & K-S & $p$-value & CvM & $p$-value & AD & $p$-value \\
\hline
MFLR & 0.1135 & 0.3123 & 0.1382 & 0.4286 & 0.8679 & 0.4341 \\
MR & 0.1393 & 0.1226 & 0.4391 & 0.0568 & 2.2503 & 0.0674 \\
MW & 0.1346 & 0.3121 & 0.1721 & 0.3291 & 1.0473 & 0.3331 \\
ME & 0.2689 & 5.994e-05 & 1.1359 & 0.0011 & 5.8381 & 0.0012 \\
MF & 0.1172 & 0.2764 & 0.1422 & 0.4149 & 1.4497 & 0.1890 \\
\hline
\end{tabular}
\end{table}

\section{Concluding remarks and future work}\label{sec7}

By employing the Modi family of distributions and using the LFR distribution as the parent model, we suggested the MLFR distribution. This distribution has the Modi exponential distribution and the Modi Rayleigh distribution as submodels. The hazard rate behavior demonstrates that the MLFR distribution is increasing, decreasing, bathtub-shaped, upside-down bathtub-shaped, or modified bathtub-shaped. Several mathematical and reliability properties are discussed, like the $r^{\text{th}}$ moment, generating function, $r^{\text{th}}$ conditional moment, mean deviations, order statistics, Rényi entropy, and reliability. The model parameters are estimated using the method of maximum likelihood, and a simulation study demonstrates that these perform well in finite samples and possess desirable large-sample properties such as asymptotic unbiasedness and consistency. Two well-known real survival data sets prove that our model gives a good fit compared to several recently introduced competing distributions, particularly those developed under the same Modi family framework. In future work, we plan to extend the MLFR distribution to multivariate and Bayesian frameworks to better accommodate complex data structures. Additional research may focus on adapting the MLFR model for censored and truncated datasets, which are common in survival analysis and reliability studies. We also aim to integrate the MLFR model with modern machine learning techniques, such as ensemble methods, neural networks, and Gaussian processes, to improve predictive performance. Finally, we will investigate appropriate goodness-of-fit tests for the MLFR distribution.

%%%%%%%%%%%%%%%%%%%%%%%%%%%%%%%%%%%%%%%%%%
 
\section*{Data Availability Statement}{No primary data are used in this paper. }

\section*{Conflicts of Interest}{The author declares no conflict of interest.}

\section*{Acknowledgments}{I thank the editor and reviewers for their comments and constructive suggestions, which have helped improve the quality of this paper.}

\section*{Funding}{This research received no external funding.}

%\clearpage
%\let\clearpage\relax

\bibliographystyle{apalike} %APA style
\bibliography{Refff}

\end{document}